\documentstyle{aipproc}

\begin{document}
\title{Neutrino Physics in a Muon Collider}

\author{Rabindra N. Mohapatra}
\address{Department of Physics, University of Maryland, College 
Park, MD-20742\thanks{Invited talk presented at the {\it Workshop on
Physics at the First Muon Collider and the Front End of a Muon 
Collider} held in Fermilab, November 6-9, 1997. Work 
supported by the National Science Foundation grant no. PHY-9421386.}}

\hfill UMD-PP-98-63

\maketitle

\begin{abstract}
A muon collider is expected to produce a high intensity neutrino beam
which is an admixture of either $\nu_{\mu}+\bar{\nu_e}$ or $\bar{\nu}_{\mu}
+\nu_e$ which can can be directed to underground detectors far away from the
source. It will not only allow a probe of the $\nu_e-\nu_{\mu}$ as well as
$\nu_{\mu}-\nu_{\tau}$ oscillations in a range of mixing angle and $\Delta 
m^2$ not probed heretofore but it will also provide information
about the mixing angle $\theta_{e\tau}$ for a wide range of $\Delta 
m^2_{\nu_e\nu_{\tau}}$ from $10^{-4}$ eV$^2$ to $10^{-1}$ eV$^2$ which
cannot be obtained from any other existing or proposed machine. One 
can also search for violations of Lorentz invariance and deviations from 
equivalence principle for neutrinos at a level which is three 
to four orders of magnitude more sensitive than possible at the 
moment. This will test for instance some unorthodox suggestions 
to understand both solar and atmospheric neutrinos using a 
single mass difference -squared between the $\nu_e$ and 
$\nu_{\mu}$. It can also test various proposed models neutrino masses
and mixings to understand existing neutrino data.
\end{abstract}

\section*{I. Introduction}

Neutrino physics is now going through a very exciting period. For the first 
time in its history, there are several hints for a nonvanishing mass
for at least two of three known neutrinos
which look very promising and credible. They come (i) from the observations
of the solar neutrinos in various experiments and their disagreement
with the predictions of the standard solar model\cite{solar}: the 
earlier experiments from Homestake, Kamiokande, SAGE and 
GALLEX\cite{expt} and the
most recent high statistics confirmation of these results by the 
super-Kamiokande experiment\cite{superK} and (ii) from the observations of 
the atmospheric
neutrinos by several previous experiments\cite{atmos,fukuda} and the most 
recent
confirmation of the earlier results by the super-Kamiokande\cite{superK}
collaboration. 
Then there is the result from the Los Alamos liquid scintillation
neutrino detector (LSND)
which gives the first laboratory evidence for the oscillation of both
$\bar{\nu}_{\mu}\rightarrow \bar{\nu}_e$\cite{LSND} as well as 
$\nu_{\mu}\rightarrow \nu_e$ type\cite{LSND2}.

Once the neutrinos have mass, they can mix with each other leading to
a rich variety of new physical phenomena, which in turn may lead to insight
into the kind of new physics responsible for such mixings and masses.
 
While at the moment detailed fits to all the above data 
considerably restrict the
nature of the masses and mixings among the three neutrino species, they 
do not fix the complete neutrino mass texture. On top of this, there are 
ambiguities that open the neutrino oscillation interpretation of the 
solar neutrino data to question. Therefore it is crucial that other 
experiments are performed not only to confirm what is known but also to gain 
complete knowledge of this basic sector of the standard model. 

Presently in planning and constuction stage are several such experiments-
MINOS and PALO VERDE to give two examples; several others which are 
beginning to provide this information are the CHORUS, NOMAD, CHOOZ.
These are variously known as long base line and short base line experiments,
which will either involve low energy electron neutrino beams from the
reactors or high energy $\nu_{\mu}$ beams from the accelerators. None of them
will have a high energy $\nu_e$ beams. 
While the disappearance of the $\nu_e$ in the reactor experiments 
one can get information about the mixing angles for a certain mass range,
a large range of mixing angles and masses of theoretical interest
remains unexplored at the moment. Specifically, information on the
$\nu_{e}-\nu_{\mu}$ mixing angle at the moment is very poor.
A similar remark also applies to the $\nu_{\mu}$ case where appearance 
experiments involving the $\nu_{\mu}$ beam
provide knowledge of the mixing of $\nu_{\mu}$ with $\nu_e$ or 
$\nu_{\tau}$ (i.e. $\theta_{\mu e}$ or $\theta_{\mu\tau}$) for some
mass range while leaving a considerable range of interest unexplored. 

The goal of this article is to explore whether the neutrino beams from a muon
collider can provide any useful information regarding the neutrino masses 
and mixings that are not already available (or will not be available once the
above mentioned experiments are completed). In other words is there a 
neutrino physics justification for the muon collider?

The muon collider will be the first place where
one can get an energetic beam of electron neutrinos. Therefore with a 
suitable long
baseline experiment, one can probe the mixing angle $\theta_{e-\tau}$ in
a completely unexplored domain. Of course needless to say that
the muon collider will also provide extensive information on 
$\theta_{e-\mu}$ or $\theta_{\mu-\tau}$ for very small mass difference 
squared as well and thus nicely complement the other experiments and may even
extend the domain of the search.

Besides with the high flux neutrino beams which are supposed to result
in a muon collider, one can also test the validity of several fundamental
laws of physics such as Lorentz invariance, CPT theorem as well
as equivalence principle and I will show that improvements by several
orders of magnitude are possible with the muon colliders in these cases.

This article is organized as follows: in section 2, the implications of 
the solar, atmospheric and the LSND experiments for neutrino masses and
mixings are briefly summarized; in section 3 a summary of the various
popular scenarios for neutrino masses and mixings are touched upon; in 
section 4, the neutrino mixings that can be probed by the muon collider
is given using different experimental scenarios and in section 5, we 
discuss possible tests of the Lorentz invariance, equivalence principle
and CPT theorem are considered.

\section*{II. Indications for nonzero neutrino mass}

\bigskip
\noindent{\bf II.a Solar neutrino deficit}
\bigskip

We will assume the explanation of the solar neutrino deficit  
in terms of the oscillation between 
the $\nu_e$ and $\nu_{x}$ where $x$ is another species of neutrino not
necessarily of muon or tau type. The oscillation can be pure vacuum 
oscillation which requires a mass difference-squared $\Delta m^2\sim 
10^{-10}$ eV$^2$ and large mixing or it could be matter enhanced 
MSW\cite{msw} type in which case the neutrino 
mass differences and mixing angles fall into one of the following 
ranges\cite{solar},
\begin{eqnarray}
  {\rm a)}&{\rm Small-angle\ MSW,\ }\Delta m^2_{ei}\sim 5\times10^{-6}   
- 10^{-5}{\rm eV}^2,
         \ \sin^22\theta_{ei}\sim7\times10^{-3},&\cr
  {\rm b)}&{\rm Large-angle\ MSW,\ }\Delta m^2_{ei}\sim9\times10^{-6}{\rm 
eV}^2,
         \ \sin^22\theta_{ei}\sim0.6. 
\end{eqnarray}
If the solar neutrinos oscillate into sterile neutrinos, the MSW effect  
is different from the $\nu_e$ to $\nu_{\mu}$ case
and the large angle solution is no more allowed. The above results are
based on the approximation that only two of the neutrino species are
involved in the oscillation.

\bigskip
\noindent{\bf II.b Atmospheric Neutrino Deficit}
\bigskip

The atmospheric $\nu_\mu$'s and $\nu_e$'s arise from the 
decays of
$\pi$'s and $K$'s and the subsequent decays of secondary muons produced 
in the
final states of the $\pi$ and $K$ decays.  In the underground experiments 
the
$\nu_\mu$ and ${\bar\nu}_\mu$ produce muons and the $\nu_e$ and 
${\bar\nu}_e$
lead to $e^\pm$.  Observations of $\mu^\pm$ and $e^\pm$ indicate a far lower
value for $\nu_\mu$ and ${\bar\nu}_\mu$ than suggested by na\"\i{}ve 
counting
arguments which imply that $N(\nu_\mu+{\bar\nu}_\mu)=2N(\nu_e+{\bar\nu}_e)$
\cite{atmos}.
The assumed oscillation in this case could a priori be between $\nu_{\mu}$
to $\nu_e$ or $\nu_{\mu}$ to $\nu_{\tau}$. However, a recent CHOOZ 
collaboration result rules out the Kamiokande allowed $\nu_{\mu}$ to $\nu_e$
mass-squared mixing region\cite{CHOOZ}. Thus we 
can assume that the oscillation of $\nu_{\mu}$ to $\nu_{\tau}$
provides the explanation of the atmospheric neutrino results. Fits to 
both the sub-GeV and multi-GeV Kamiokande data require that\cite{fukuda}
\begin{eqnarray}
\Delta m^2_{\mu \tau}\approx0.025{\ \rm to\ 0.005\ eV}^2,\
sin^22\theta_{\mu \tau}\approx .6~~to~~1.
\end{eqnarray}

The most recent Super-Kamiokande data has confirmed the deficit in both 
the  
sub-GeV and the multi-GeV data.
Also there is now evidence for zenith angle dependence in the multi-Gev data
which according to preliminary analysis\cite{superK} would indicate a
similar mass range as above for maximal mixing angle.
\bigskip

\noindent{\bf II.c Results from the LSND experiment}  
\bigskip

The LSND collaboration first reported seeing indications for   
$\bar{\nu}_{\mu}$ to $\bar{\nu}_e$ oscillation using the liquid 
scintillation
detector at Los Alamos in 1996\cite{LSND}.
Their results in conjunction with the negative results by the E776 group
and the Bugey reactor data imply a mass difference
squared between the $\nu_e$ and the $\nu_{\mu}$ lying between
\begin{eqnarray}
0.27~eV^2\leq \Delta m^2 \leq 10~eV^2
\end{eqnarray}
with a mixing angle $\theta_{e\mu}\sim .05 -.1$.
The region for $\Delta m^2$ above 10 eV$^2$ has been ruled out both
by the recent CCFR data and the NOMAD data\cite{zuber}. In a recent
paper \cite{LSND2}, LSND group has reported preliminary evidence for the
$\nu_{\mu}-\nu_e$ oscillation with mass difference squares and mixings in 
the similar range as above. 

\bigskip

\noindent{\bf II.d Hot dark matter of the universe}
\bigskip

There is increasing evidence that more than 90\% of the mass in the 
universe 
must be detectable so far only by its gravitational effects.  This dark 
matter
is likely to be a mix of $\sim20$\% of particles which were relativistic 
at the
time of freeze-out from equilibrium in the early universe (hot dark 
matter) and
$\sim70$\% of particles which were non-relativistic (cold dark matter).  
Such a mixture gives a very good fit to all available cosmological 
data\cite{primack}.
This interpretation is however by no means unique and it has been 
claimed that an equally good fit to the power spectrum can be obtained by
a pure CDM model with a tilted spectrum\cite{sarkar}.
  
If however, the mixed dark matter picture is adopted, a very plausible
candidate for hot dark matter is one or more species of neutrinos with total
mass of $\Sigma_im_{\nu_i}=93h^2F_H\Omega=4.8$ eV, if $h=0.5$ (the Hubble 
constant
in units of 100 km$\cdot$s$^{-1}\cdot$Mpc$^{-1}$), $F_H=0.2$ (the 
fraction   
of dark matter which is hot), and $\Omega=1$ (the ratio of density of the
universe to closure density).

It is usually assumed that the $\nu_\tau$ would supply the hot dark 
matter. 
However,
if the atmospheric $\nu_\mu$ deficit is due to $\nu_\mu\to\nu_\tau$, the
$\nu_\tau$ alone cannot be the hot dark matter, since the $\nu_\mu$ and
$\nu_\tau$ need to be closer to each other in mass.
It is interesting that instead
of a single $\sim 4.8$ eV neutrino, sharing that $\sim 4.8$ eV
between two or among
three neutrino species provides a better fit to the universe structure 
and  
particularly a better understanding of the variation of matter density with
distance scale\cite{HDM}.

\bigskip
\noindent{\bf II.e: Neutrinoless double beta decay constraints}
\bigskip

Finally, let us note the very stringent constraints on neutrino masses now
implied by the neutrinoless double beta decay searches.
 The Heidelberg-Moscow $^{76}$Ge
experiment\cite{klap} has provided the most stringent upper limits
on the effective Majorana mass of the neutrino: $<m_{\nu}>\leq .47$ eV
where $<m_{\nu_e}>=\Sigma_i U^2_{ie} m_{\nu_i}$. This is beginning to put
very strong constraints on model building. For instance, it has recently been
noted\cite{kim} that the CHOOZ and Bugey\cite{bugey} results already imply
that $|<m>|\leq 3\times 10^{-2}$ eV. Thus any signal for neutrinoless
double with $<m>$ above 0.1 eV would be evidence against a hierarchical
neutrino mass pattern. Similarly, one can infer from the LSND data that
one must have $<m>\geq 4\times 10^{-3}$ eV assuming that $\theta_{e\tau}$
is small (or at least it does not precisely cancel this contribution).
Thus high precision double beta searches are extremely important to a
complete understanding of the neutrino masses.

\section*{ III. Neutrino mass textures implied by data}

	In order to discuss the implications of the above data for the 
neutrino mass pattern,
we will assume that all the neutrinos are Majorana particles, since
it is easier to understand the smallness of Majorana masses of neutrinos
within the framework of grand unified theories. We will then proceed
by assuming that the solar and the atmospheric neutrino data are the two
core items that appear as the most secure indications of neutrino 
oscillation and study their significance for neutrino masses. We
will then add the HDM and the LSND results and see their implications.

\bigskip
\begin{center}
\underline{\bf : Including only solar and the atmospheric data:}
\end{center}
\bigskip

Since we only have constraints on the mass difference squares from the
solar and the atmospheric data, we can have a ``staircase'' pattern
with $m_{\nu_e}\ll m_{\nu_{\mu}}\simeq\sqrt{ \Delta m^2_{solar}}$ and 
$m_{\nu_{\mu}}\ll m_{\nu_{\tau}}\simeq \sqrt{\Delta m^2_{atmos}}$
or a degenerate pattern\cite{caldwell,other} where all masses are nearly 
equal with appropriate
mass differences. The latter is mandatory if one wants to explain the
HDM picture of the universe. As far as the mixing angles go, the
$\theta_{\mu\tau}$ is always maximal (i.e. near $\pi/4$) whereas 
$\theta_{e\mu}$ is either maximal (for vacuum oscillation or large angle
MSW) or few percent (for small angle MSW). Several theoretical suggestions
are now given.

\bigskip


\noindent{\bf III.a Model A: Maximal mixing scheme}
\bigskip

In this scheme\cite{nussinov}, the mixing matrix has the form:
\begin{eqnarray}
U_{\nu}=\frac{1}{\sqrt{3}}\left(\begin{array}{ccc}
1 & 1 & 1 \\
1 & \omega & \omega^2 \\
1 &\omega^2 & \omega \end{array}
\right)
\end{eqnarray}
where $\omega = e^{2\pi i/3}$ 
This scheme becomes essential if the neutrino masses are degenerate and
if the limits on the neutrino mass from neutrinoless double beta decay
keps going down\cite{mohnus} since the leading term in the $<m_{\nu_e}>$
cancels for this choice of mixing angles. This mixing angle pattern can
be shown to fit both the solar and atmospheric neutrino 
data\cite{mohnus,perkins} if one assumes vacuum oscillation solution to the
solar neutrino problem.

\bigskip
\noindent{\bf III.c Model B: Democratic mixing among neutrinos}
\bigskip

This model for mixings is based on the idea that the neutrino mass matrix
may satisfy an approximate permutation symmetry among the three 
generations\cite{fritzsch} and also can fit the solar and atmospheric 
neutrino data and has a mixing matrix of the following form:
\begin{eqnarray}
U_{\nu}=\left(\begin{array}{ccc}
\frac{1}{\sqrt{2}} & -\frac{1}{\sqrt{2}} & 0\\
\frac{1}{\sqrt{6}} &\frac{1}{\sqrt{6}} &-\frac{2}{\sqrt{6}} \\
\frac{1}{\sqrt{3}} &\frac{1}{\sqrt{3}} &\frac{1}{\sqrt{3}}
\end{array}\right)
\end{eqnarray}

This model also can support a degenerate mass pattern consistent with 
neutrinoless double beta decay. The difference between this and the 
maximal mixing pattern is that the $\theta_{e\tau}$ values are very
different.

There are several other schemes based on attractive theoretical
assumptions that lead to three neutrino mixing patterns\cite{koide} that can 
fit both solar and atmospheric data. 

\bigskip

\begin{center}
\underline{\bf : Accomodating solar, atmospheric and the LSND data:}
\end{center}
\bigskip

The LSND result have two important implications for our discussion:
first, their result implies oscillation from $\bar{nu}_{\mu}$ to 
$\bar{\nu}_e$ i.e. unlike the solar and atmospheric data, the final
state is not a matter of speculation but observation; secondly the
$\Delta m^2_{e\mu}$ that fits data is between $.2$ eV$^2$ to about 
10 eV$^2$, which is very different from the ranges derived from simple
interpretations of the data as noted above. Before the latest results
from super-Kamiokande experiment came out, two interesting neutrino mass 
schemes were proposed which seemed in accord (though rather marginally)
with the previous Kamiokande data. The basic idea in these papers was
the following: three experiments are sensitive to three mass difference 
squares; however with three neutrinos there are only two possible 
$\Delta m^2$'s. Therefore a three neutrino scheme can only fit data
if two of the experimentally determined $\Delta m^2$'s turn out to be
equal. The two models described below essentially exploit these two
possibilities.

 \bigskip 
\noindent{\bf III.c Model C: Cardall-Fuller scheme}
\bigskip

This scheme\cite{fuller2} assumes that $\Delta m^2_{LSND} \simeq \Delta 
m^2_{atmos}$ and that the $\nu_{\mu}-\nu_e$
oscillation observed at Los Alamos
is an indirect oscillation\cite{babu} which proceeds
as $\nu_e$ to $\nu_{\tau}$ to $\nu_{\mu}$.
To accomodate the LSND results in this picture assumes
the LSND $\Delta m^2$
to be around $.3$ eV$^2$. Since the
solar neutrino puzzle requires that $\Delta
m^2_{e-\mu}\simeq 10^{-5}$
eV$^2$, this scenario implies that we
must have $\Delta^2_{\mu-\tau}$ be $\approx .3~eV^2$.
so that the LSND neutrino oscillation frequency
is determined by $\nu_e$-$\nu_{\tau}$ mass difference.
Secondly, for the amplitude of indirect oscillation to
be compatible with
observations, the $\nu_{e}-\nu_{\tau}$ mixing angle
should be nonnegligible
(say $\sim .1-.2$). The main problem
for this scenario comes from the atmospheric neutrino
data,
since the original analysis of the Kamiokande sub-GeV 
and
the multi-GeV data by the Kamiokande group excludes
$\Delta m^2\geq .1~eV^2$
at 90\% confidence level (c.l.). The analysis of the
the atmospheric neutrino
data from Super-Kamiokande will therefore provide
crucial test of this model. Preliminary analysis of the
super-Kamiokande data (which has a clear evidence for zenith angle 
dependence) appears to contradict this scenario. Furthermore, if we want to 
fit the HDM picture into   this model, one must have
$m_{\nu_e}\simeq 1.6$ eV.
While at its face this value
may be in conflict with the neutrinoless double beta decay results
\cite{klap}, one can hide under the uncertainties of
nuclear matrix element
calculations which typically could be as much as a
factor of 2-3.
As the precision in $\beta\beta_{0\nu}$ search improves
further (say
to the level of $0.1$ eV), nuclear matrix element
uncertainties cannot be invoked to save the model anymore.

\bigskip

\noindent{\bf III.e Model D: Acker-Pakvasa scheme}
\bigskip

The second three neutrino mass texture\cite{pakvasa}
also uses indirect oscillation to explain the LSND data but
makes the assumption that $\Delta m^2_{solar}\simeq \Delta m^2_{atmos}$
and assumed that the atmospheric neutrino oscillation involves 
$\nu_{\mu}$ to $\nu_e$ oscillation, which looks implausible in view of the
latest CHOOZ data. In any case, they choose
$\Delta m^2_{e\mu}\simeq 10^{-2}$ eV$^2$, $\Delta
m^2_{e\tau}\simeq
\Delta m^2_{\mu\tau}\simeq 1-2~eV^2$. It is easy to
see that in this case
the general three neutrino oscillation formula for
$P_{ee}$ becomes energy independent if $L$
is chosen to correspond
to the distance of the earth from the Sun. 
It was shown in Ref.\cite{pakvasa} that if one 
reduces the $^8$B production
in the center of the Sun, one can fit all 
solar neutrino observations despite the energy independence of the 
oscillations.
It has been pointed out
that already in the present data, there is 
evidence for energy
dependence\cite{krastev} disfavoring this 
scheme. Therefore this can also be
tested by the Super-Kamiokande observations. 

To complete this model, we give a typical mixing matrix that characterizes
this model:
\begin{eqnarray}
U_{\nu} = \left( \begin{array}{ccc}
.700 & .700 & .14 \\
-.714 & .689 & .124 \\
-.010 & -.187 & .982 \end{array} \right)
\end{eqnarray}
Note the large value of $\Theta_{e\tau}$.

\bigskip

\noindent{\bf III.e Model E: The case for a sterile neutrino}
\bigskip

The case for a sterile neutrino is made clear by noting the difficulty
of fitting the solar, atmospheric and the LSND data with three neutrinos as
exemplified by the models C and D.
The main obstacle, as we saw, comes from the conflict between the
LSND data and the MSW resolution of the solar neutrino data.

The general picture for the case of sterile neutrino is as
follows\cite{caldwell,valle,carlo}: the solar neutrino puzzle is explained 
by the
$\nu_e-\nu_s$ oscillation; atmospheric neutrino data would be explained by
the $\nu_{\mu}-\nu_{\tau}$ oscillation. The LSND data would set the
overall scale for the masses of $\nu_{\mu}$ and $\nu_{\tau}$ (which are
nearly degenerate)  and if this
scale is around 2 to 3 eV  as is allowed by the data\cite{LSND}, then
the $\nu_{\mu,\tau}$ would constitute the hot dark matter of the universe.
The simplest (though by no means unique) mass matrix in this case would be
 in the basis ($\nu_s$, $\nu_e$, $\nu_\mu$, $\nu_\tau$),
\begin{eqnarray}   
M=\pmatrix{\mu_1&\mu_3& 0& 0\cr
             \mu_3& 0 & 0 &\epsilon \cr
             0& 0 &\delta & m\cr
             0 & \epsilon & m &\delta\cr}.
\end{eqnarray}
Solar neutrino data requires $\mu_3\ll \mu_1
\simeq 10^{-3}~eV$ and $\epsilon \simeq .05 m $.
The $\epsilon $ term is responsible for the $\nu_e-\nu_{\mu}$ oscillation.  
Clearly the crucial test of the sterile neutrino scenario will   
come when SNO collaboration obtains their results for neutral current
scattering of solar neutrinos. One would expect that $\Phi_{CC}=\Phi_{NC}$
if the $\nu_e$ oscillation to $\nu_s$ is responsible for the solar neutrino
deficit. There should be no signal in $\beta\beta_{0\nu}$ search. Precision
measurement of the energy distribution in
charged current scattering of solar neutrinos at Super-Kamiokande can also
shed light on this issue.

\bigskip

\section*{IV. Muon collider 
for studying the neutrino masses and mixings}

The muon collider is expected to produce a high luminosity                   
beam of neutrinos which is an admixture of either 
$\nu_{\mu}+\bar{\nu}_e$ or their antiparticles. Thus in some 
sense this is a ``controlled atmospheric neutrino'' beam. The
energy of the neutrinos is expected to range from 10 GeV to 100 
GeV. In our discussion we will entertain the possibility of two 
kinds of long base line experiments\cite{geer} with beam directed 
to either Gran Sasso or Soudan mine with distances respectively of
$\sim 10^4$ or 750 kilometers. Recall that the oscillation formula
\begin{eqnarray}
P(\nu_e\rightarrow \nu_{\mu}) = sin^22\theta sin^2\frac{1.27 
\Delta m^2 L}{E}
\end{eqnarray}
where E is in GeV and L is in kilometers. With $7.5\times 10^{20}$
$\mu^{\pm}$ per year, one can have of the order of $10^{20}$ neutrinos/year
\cite{geer}. Geer has calculated the charge current event rate for such
particles at a 10 kiloton detector located at Gran Sasso as well as at 
Soudan. He finds that one can expect thousands of charged current events
per year.

To discuss its utility in studying neutrino masses and mixings, note that
for a 10 GeV neutrino beam and a distance of $10^4$ kilometers, one could 
probe $\Delta m^2$ down to $10^{-5}$ eV$^2$ for maximal mixing 
if we take 10 
events per year to get a signal  and for $\Delta m^2\geq 10^{-3}$ eV$^2$,
one could probe mixings $sin^22\theta$ down to $10^{-4}$ or so. Thus, one can
not only explore $\theta_{e\tau}$ in a totally unexplored region of 
parameters but also  considerably extend our knowledge of the $\theta_{e\mu}$
as well as $\theta_{\mu\tau}$ into a domain further than what MINOS or COSMOS
can accomplish. As a comparision, note that at present, $\theta_{e\tau}$ has
an upper bound of about $.14$ from the Fermilab experiment E531 for 
$\Delta m^2\geq 10~ eV^2$, which is a very weak bound compared to 
the other two mixing angles. The main reason being that there does not exist
any accelerator source of high energy $\nu_e$'s and muon collider will be 
the first one to provide one such source.

\bigskip

\section*{IV. Testing Lorentz and CPT invariance and equivalence principle}

Lorentz invariance, CPT invariance (which is a consequence of Lorentz
invariance and locality in Quantum Field theories) and the principle
of general covariance are some of the fundamental pillars on which the
present day theoretical physics rests. While few would doubt that there
is any deviation from these principles, science has to be based on 
experimentally tested ideas.  It is therefore important to look for ways
to test the validity of these principles. In order to make the tests 
quantitative, a framework that has
some parameters that characterize the departures from the exactness of these 
principles is useful.
Such frameworks have recently been discussed and I summarize them below and
point out how a muon collider can be useful.

\bigskip
\noindent{\bf IV.a  Lorentz invariance}
\bigskip

It was pointed out recently by Coleman and Glashow\cite{coleman}
that one way to parameterize a departute from Lorentz invariance for massless
particles such is to write
\begin{eqnarray}
E_i=p(c+\delta c_i)
\end{eqnarray}
Applying this to neutrinos, one gets for the energy difference between two
 eigenstates into which the weak eigenstate resolves as 
$E_1-E_2=E(\delta 
c_1-\delta c_2)\equiv E\delta v$. One can then write the oscillation 
probability of say $\nu_e$ to $\nu_{\mu}$ to be 
\begin{eqnarray}
P(\nu_e\rightarrow \nu_{\mu})= sin^22\theta_v sin^2\frac{\delta vEL}{2}
\end{eqnarray}
 
The energy dependence of the $P(\nu_e\rightarrow\nu_{\mu})$ in Eq. 11
is clearly very different from the case of mass oscillation where it goes like
$L/E$. Therefore longer the base line and higher the energy, the more
precise the test of Lorentz invariance. Present limits on the $\delta v$
from various oscillation experiments is $\delta v\leq 10^{-21}$. In this 
case we must choose the neutrino energy from the muon collider to be as high 
as possible. Again taking $E\simeq 100$ GeV and $L= 10^{4}$ Km, this limit
can be improved to $\delta v\leq 10^{-26}$ which is an improvement of 
some five orders of magnitude.

\bigskip
\noindent{\bf IV.b  CPT for neutrinos}
\bigskip

A simple CPT violating combination of oscillation probabilities for the
$\nu_e-\nu_{\mu}$ system is given by $P(\nu_{\mu}\rightarrow{\nu}_e)-
P(\bar{\nu}_e\rightarrow \bar{\nu}_{\mu})$ as is very easily checked. Note 
that as mentioned before the neutrino beam in a muon collider consists of
$\nu_{\mu}$ and $\bar{\nu}_e$. Let us assume that $N_{\mu}=N_{\bar{\nu}_e}$
(although the energy spectra in general will be different) for simplicity.
Then without CPT violation but with $\nu_e-\nu_{\mu}$ oscillation, one would 
expect in the detector $N_{e^-}
=N_{\mu^+}$. Thus any deviation from this equality would be a test of CPT
violation. This result is independent of any specific underlying model
for CPT violation.

\bigskip

\noindent{\bf IV.c Violations of equivalence principle }
\bigskip

It has been pointed out by Halprin and Leung\cite{leung}
that violations of equivalence principle can also lead to neutrino
oscillation phenomena. To see this, let us parameterize the metric as
\begin{eqnarray}
Metric= g_{\alpha\beta}+2\gamma_i \phi \delta_{\alpha\beta}
\end{eqnarray}
where $\phi$ is the gravitational potential. The second term is absent in 
Einstein's theory and characterizes the departure from the equivalence 
principle. The energy-momentum relation now looks as follows: 
\begin{eqnarray} E^2(1+2\gamma_j\phi)=p^2(1-2\gamma_j\phi)
\end{eqnarray}
This can be cast in the language discussed in connection with violation of
Lorentz invariance identifying $\delta v\equiv 2(\gamma_1-\gamma_2)\phi$.
Translating our earlier discussion then we can conclude that in a muon 
collider, one can probe $2\Delta \gamma \phi$ down to the level of $10^{-26}$
as before. Since this experiment will be done in the solar system, the 
value of $\phi\simeq 10^{-6}$, it will test for violation of equivalence 
principle down to the level of $10^{-20}$. Note the present long range
force experiments test this principle down to the level of $10^{-12}$ or 
so.

More importantly, Halprin et al have made the unconventional suggestion that 
perhaps one 
could use this phenomenon to explain solar and atmospheric neutrino puzzle
by setting $2\Delta\gamma\phi\simeq 10^{-21}$ using only 
$\nu_{\mu}-\nu_e$ oscillation. A 
muon collider could therefore provide a clean test of this hypothesis.

In conclusion, we find that the neutrino beams from the muon collider
can provide extremely useful insight into the world of neutrino masses and
mixings, specifically it can probe the $\nu_e-\nu_{\tau}$ mixing angle in a 
domain of parameters that is beyond the range of any proposed experiment.
This will allow us to test several three neutrino mixing schemes such
as the maximal mixing scheme and the SO(10) scheme.
Muon collider can also extend the domain of validity of some of the 
fundamental 
laws govorning the physical phenomena such as the equivalence principle,
CPT theorem and Lorentz invariance.

\begin{center}
{\bf Acknowldgement}
\end{center}

I would like to thank Boris Kayser for the invitation to join the muon
collider working group on neutrinos and many stimulating discussions on 
the topics discussed in this article. I also like to thank G. 
Sullivan for discussions on the super-Kamiokande results for atmospheric 
neutrinos and B. Dutta and S. Nandi for discussions on the four neutrino
scheme.

\end{document}